\UseRawInputEncoding
\documentclass[11pt,aps,amssymb,prd,twocolumn,a4paper,nofootinbib,tightenlines]{revtex4-2}
\usepackage{fullpage}
\usepackage{amsfonts}
\usepackage{amsmath}
\usepackage{slashed}
\usepackage{amssymb}
\usepackage{graphicx}
\usepackage{makeidx}
\usepackage{cancel}
\usepackage{epic}
\usepackage{eepic}
\usepackage{epsfig}
\usepackage{latexsym}
\usepackage[dvipsnames]{xcolor}
\usepackage{float}
\usepackage{multirow}
\usepackage[export]{adjustbox}
\usepackage{xurl,hyperref}
\usepackage{enumitem}
\hypersetup{colorlinks=true,citecolor=red,linkcolor=NavyBlue,urlcolor=NavyBlue}
\usepackage[utf8]{inputenc}
\usepackage[caption=false]{subfig}
 
\usepackage{natbib}
\usepackage{relsize}
\usepackage[left=1.8 cm,right=1.5 cm,top=2 cm,bottom=2 cm]{geometry}
\usepackage{mathptmx}
\begin{document}
%\relscale{1.05}
\captionsetup[subfigure]{labelformat=empty}

\title{The 248 keV LZ Recoil: A Possible Hint of Non-SM-Like Quark Yukawa Couplings with a Scalar-Portal Dark Matter}

\author{Bibhabasu De}
\email{bibhabasude@gmail.com}
\affiliation{Department of Physics, The ICFAI University Tripura, Kamalghat-799210, India}

\date{\today}

\begin{abstract}
\noindent
The recent observation of an isolated nuclear recoil event at an energy of $248\pm 23\,{\rm (stat)}\pm 23\,{\rm (sys)}$ keV, as reported by the LUX-ZEPLIN~(LZ) collaboration, can be an intriguing signature for Beyond the Standard Model~(BSM) possibilities, particularly for dark matter~(DM). This paper explores a scenario in which the observed recoil could simultaneously signal inelastic DM and a flavor-specific {\it New Physics}~(NP) interaction in the quark sector. Considering a minimal extension of the Standard Model~(SM) with two closely degenerate $Z_2$-odd vector-like fermions~(VLFs), interacting with the SM via a scalar portal, the analysis shows that a perfect agreement with the observed 248 keV nuclear recoil can be achieved if the quark Yukawa couplings are allowed to deviate from their SM values. The underlying quarkophilic NP can originate from a dimension-6 effective operator at a NP scale $\Lambda\leq \mathcal{O}(10)$ TeV. Future collider searches can be crucial to test/falsify the proposal.
\end{abstract}
	
\maketitle	

\section{Introduction}
\noindent
The LZ collaboration, with an exposure of 2.84 tonne-years, has recently reported one event~(designated as LZ230616) at a recoil energy, $E_R=248\pm 23\,{\rm (stat)}\pm 23\,{\rm (sys)}$ keV~\cite{LZ:2026axp} when the energy window was extended up to 270 keV. The observation corresponds to a maximum local significance of $3.4\sigma$, while after accounting for the look-elsewhere effects, it is reduced to a global significance of $2.6\sigma$. However, a single event with a significance far below the $5\sigma$ value can't result in a concrete conclusion. Nevertheless, it is phenomenologically vital to identify the theoretical constructions that can produce such an isolated high-energy signal with a low background expectation. Though the elastic scattering of weakly interacting massive particles~(WIMPs) with masses above 75 GeV can produce 248 keV nuclear recoil energy at a Xe-based detector, the absence of any low-energy signal can't be explained with the conventional elastic scatterings\,\footnote{Ref.~\cite{Bose:2026szs} has imposed stringent constraints on the elastic DM explanations of LZ230616 using the solar capture of WIMPs.}. Moreover, due to suppression from the velocity distribution and nuclear form factor, elastic scattering spectra of canonical WIMPs fall by four to six orders of magnitude within the energy range $E_R\sim\mathcal{O}(10-100)$ keV. However, inelastic scattering can be a viable explanation for LZ230616, as the associated kinematics impose a threshold to intiate the recoil. Several theoretically motivated DM models have already appeared to explain the observed LZ event~\cite{Su:2026rwz,Freese:2026sga,Wu:2026nhi,Lou:2026idn,Yin:2026jnn,DiMauro:2026ldr,Yamashita:2026ump,Chattopadhyay:2026ryw,Du:2026guj,Rodd:2026tyn,McCabe:2026crm,Unwin:2026rdp,Dent:2026bji,deLima:2026shq,Gu:2026vto,Lee:2026wof,Baer:2026fpy,Wang:2026ytg,Kotlarski:2026pep,Liang:2026coz,DiMauro:2026dqp,Alhazmi:2026efz,Okada:2026eol,Ahmed:2026qjg,Du:2026lpa,Bandyopadhyay:2026gjw,Kannike:2026qyl,Bisal:2026khf,Cheung:2026byg,Yuan:2026djt,Elahi:2026vlm,Zhu:2026dag,Aghaie:2026vsu,Lee:2026xxh,Lee:2026jxl,Khan:2026nwp,Langhoff:2026ujr,Chatterjee:2026scv,He:2026hqz,Fan:2026hzw,Qi:2026vyp,Kumar:2026lgi,Frolovsky:2026tvq,Heikinheimo:2026kwp,Borah:2026ris,Barman:2026omh,Das:2026buc,He:2026idw,Lian:2026hpm,Xing:2026civ,Nagata:2026pbj,Mahapatra:2026glu,Palmisano:2026kuj} whereas, a few alternative works have also emarged, showing that LZ230616 may have a non-DM origin~\cite{Jeesun:2026vzo,Chattaraj:2026fxn,Lee:2026zbr,Okada:2026upm,Cabo-Almeida:2026uqw}. Most of the models, particularly the ones with a Higgsino DM, realize the inelastic scattering through a vector-portal framework~(i.e., either with the SM $Z$ boson or a dark photon) where the experimental constraints are comparatively stronger~\cite{Bose:2026ndd,Nguyen:2026lui,Ghosh:2026txe,Barducci:2021egn,Bauer:2018onh,Caputo:2025avc}. Further, the axion-portal DMs also suffer from several experimental bounds~\cite{Freytsis:2009ct,Gninenko:2026mgn,Kelly:2020dda} which can be easily evaded for a generic scalar-portal DM. Refs.~\cite{Das:2026uyy,DiMauro:2026ymt} have discussed the inelastic scattering of a scalar-portal DM in the context of LZ230616.

The present paper has identified a significant possibility in which LZ230616 can simultaneously indicate an inelastic DM-nucleus scattering as well as the presence of a quarkophilic NP. The basic framework follows from a non-supersymmetric economical extension of the SM with two closely degenerate SM-singlet VLFs and an SM-singlet scalar. The SM gauge group has also been extended with a local $U(1)^\prime$ and $Z_2$ symmetries. The VLFs and the scalar are non-trivially charged under the $U(1)^\prime$, while all the SM fields transform as singlets. With a soft $U(1)^\prime$-breaking term, a predominantly off-diagonal coupling can be maintained between the VLFs. Note that the VLFs are protected by the exact $Z_2$ symmetry, making the lightest one a viable DM candidate. Further, the same soft $U(1)^\prime$-breaking term plays a crucial role in inducing a non-negligible mixing between the BSM scalar and the SM Higgs, leading to two physical scalar fields acting as portals between the VLFs and the SM sector. Thus, the model fulfills the minimum requirement of a quark Yukawa~($y_q$) depedent inelastic DM scattering. With a mass splitting of $\mathcal{O}(100)$ keV between the two VLFs, the analysis shows that one can have an excellent agreement with LZ230616 if the quark Yukawa coupling(s) can deviate from their SM values, particularly if $y_q<0$. Note that in the presence of a flavor-specific dimension-6 effective operator at a NP scale $\Lambda$, the linearity between $y_q$ and the quark masses can be broken. Moreover, the experiments being less sensitive to light quark Yukawa couplings, the present model considers a case where only the down quark Yukawa coupling can assume a negative value. With a specific choice corresponding to a significant isospin violation, one event at a recoil energy of 248 keV can be observed for a $\mathcal{O}(1)$ TeV thermal DM.    

The rest of the paper has been structured as follows. Sec.~\ref{sec:model} defines the BSM interactions and the soft breaking of $U(1)^\prime$, resulting in scalar-mediated off-diagonal interactions between the two VLFs. In Sec.~\ref{sec:relic}, the parameter space has been tested for DM relic density through thermal freeze-out. Sec.~\ref{sec:scat} discusses the computation of the differential event rate for an endothermic inelastic scattering between the VLFs for the two cases: $y_q=y_q^{\rm SM}$, and $y_d<0$. Finally, the work has been concluded in Sec.~\ref{sec:conc}. 
\section{The Model}
\label{sec:model}
\noindent
The paper considers a simple extension of the SM gauge group~($\mathcal{G}_{\rm SM}$) with a new Abelian gauge symmetry, $U(1)^\prime$. Two nearly degenerate SM-singlet VLFs, $\chi_1$ and $\chi_2$, and an SM-singlet complex scalar $\Phi$ are added to the particle spectrum with non-trivial $U(1)^\prime$ charges. Moreover, a discrete $Z_2$ symmetry is imposed to stabilize the VLFs once the $U(1)^\prime$ is broken through the $Z_2$-even scalar $\Phi$. Thus, in the subsequent discussions, $\chi_1$ and $\chi_2$ will be frequently called dark fermions~(DFs), with $\chi_1$ standing for the lightest $Z_2$-odd state. Note that all the SM fields transform as $Z_2$-even singlets under the $U(1)^\prime$. Therefore, the BSM interactions can be cast as, 
\begin{align}
&\mathcal{L}_{\rm BSM}=\sum\limits_{j=1,\,2}\overline{\chi}_j(i\slashed{D}-m_j)\chi_j+(D^\alpha\Phi)^*(D_\alpha\Phi)\nonumber\\
&\qquad-\tilde{M}^2\left(\Phi^*\Phi\right)-\lambda_\Phi\left(\Phi^*\Phi\right)^2-\lambda_{H\Phi}(H^\dagger H)\left(\Phi^*\Phi\right)\nonumber\\
&\qquad-\left[Y\overline{\chi}_1\Phi\chi_2+{\rm h.c.}\right]\,.
\label{eq:BSM}
\end{align}
Here $D_\alpha=\partial_\alpha-iQ^\prime_k g^\prime Z_\alpha^\prime$ denotes the covariant derivative for the SM-singlet fields. $Q^\prime_k$ is the $U(1)^\prime$ charge of the particle $k$, $g^\prime$ is the $U(1)^\prime$ gauge coupling, and $Z^\prime$ is the corresponding gauge boson. However, the SM fields being singlets under $U(1)^\prime$, $Z^\prime$-mediated interactions are confined within the BSM sector only. Note that the Yukawa term $\overline{\chi}_1\Phi\chi_2$ follows from a particular $U(1)^\prime$ charge assignment, i.e., $Q_\Phi^\prime+Q^\prime_{\chi_2}=Q^\prime_{\chi_1}$. The same charge assignment also prohibits any other renormalizable Yukawa interaction between the DFs and $\Phi$. Thus, Eq.~\eqref{eq:BSM} represents a $\mathcal{G}_{\rm SM}\otimes U(1)^\prime\otimes Z_2$-allowed theory. Evidently, in the considered framework, DFs are completely secluded from the SM. Though in general, there can be kinetic mixing between the $U(1)^\prime$ and hypercharge gauge bosons, a tree-level term has been neglected in the present paper. 

As we are particularly interested in inelastic scatterings of the DFs, let's consider a soft $U(1)^\prime$ symmetry-breaking term as follows.
\begin{align}
\mathbb{V}_{\rm Soft}=\frac{\mu}{2}(H^\dagger H)\left(\Phi+\Phi^*\right)\equiv \frac{\mu}{\sqrt{2}}(H^\dagger H)\phi\,.
\label{eq:soft}
\end{align}
where, $\phi=\sqrt{2}\times{\rm Re}[\Phi]$ and $\mu$ is a mass dimensional coupling. After electroweak symmetry breaking~(EWSB), the SM-Higgs appears as,
\begin{align}
H=\frac{1}{\sqrt{2}}\left(\begin{array}{c}
0\\
h + v
\end{array}\right)\,,
\end{align}
where, $v=246.22$ GeV defines the electroweak vacuum expectation value~(VEV). Though for $\mu\neq 0$, the minimum of the scalar potential shifts from $\langle\Phi\rangle=0$, the deviation is negligible for $\mu\ll v$. However, in the presence of $\mathbb{V}_{\rm Soft}$, EWSB results in a mixing between $\phi$ and $h$. Therefore, in the basis of $(\phi\quad h)^T$, the scalar mass matrix can be defined as,
\begin{align}
\mathbf{M}^2_{\rm S}=\left(\begin{array}{c c}
M_\phi^2 & \mu v/2\sqrt{2}\\
\mu v/2\sqrt{2} & M_h^2
\end{array}\right)
\end{align}
where $M_h$ and $M_\phi=\sqrt{\tilde{M}^2+\lambda_{H\Phi}v^2/2}$ denote the masses of $h$ and $\phi$, respectively. The scalar eigenbasis can be defined as,
\begin{align}
\left(\begin{array}{c}
h_1\\
h_2
\end{array}\right)=\left(\begin{array}{c c}
\cos\theta & \sin\theta\\
-\sin\theta & \cos\theta
\end{array}\right)\left(\begin{array}{c}
\phi\\
h
\end{array}\right)\,,
\end{align}
with
\begin{align}
M_{h_1}^2=M_\phi^2\cos^2\theta+M_h^2\sin^2\theta+\frac{\mu v}{2\sqrt{2}}\sin 2\theta\,,\nonumber\\
M_{h_2}^2=M_h^2\cos^2\theta+M_\phi^2\sin^2\theta-\frac{\mu v}{2\sqrt{2}}\sin 2\theta\,,
\end{align}
representing the masses of the physical scalar fields. The mixing angle $\theta$ is given by,
\begin{align}
\theta=\frac{1}{2}\tan^{-1}\left(\frac{\mu v/\sqrt{2}}{M_\phi^2-M_h^2}\right)\,.
\end{align}
The $h_2$ can be identified as the SM Higgs with $M_{h_2}=125$ GeV, while $h_1$ represents a lighter SM-singlet-like scalar state. Thus, the present analysis follows a mass hierarchy, $M_{h_1}<M_{h_2}$. Further, following the collider constraints~\cite{Bosse:2026bdk}, $\sin\theta=0.05$ can be fixed for all the ensuing computations.
In the physical basis, the Yukawa interaction between $\chi_{1,\,2}$ and $\Phi$ can be recast as,
\begin{align}
-\mathcal{L}_{\rm Yukawa}^{\,\chi}=Y_\chi\overline{\chi}_1\left(h_1\cos\theta-h_2\sin\theta\right)\chi_2+{\rm h.c.}\,,
\label{eq:Yx}
\end{align}
where $Y_\chi=Y/\sqrt{2}$ and it can be assumed to be real without any loss of generality. Thus, the scalars $h_{1,\,2}$ act as the portals between the DM and the SM sector. Note that the soft $U(1)^\prime$ breaking is crucial to maintain a strictly off-diagonal coupling between the DFs. In principle, the tiny induced VEV of $\Phi$ can generate a mixing between $\chi_1$ and $\chi_2$ leading to the diagonal couplings as well in the physical basis. However, as $\langle\Phi\rangle\sim\mathcal{O}(\mu)$~[numerically verified], the mixing between the DFs can be neglected safely in comparison to the associated mass scale, i.e., $m_{1,\,2}\geq\mathcal{O}(100)$ GeV. It's worth emphasizing that, in the case of the DFs, the mixing is proportional to $Y_\chi\langle\Phi\rangle\sim Y_\chi\mu$, while for the scalars, the mixing term goes as $\mu v$, making it $\mathcal{O}(v/Y_\chi)$ larger than the fermionic counterpart.
\section{Relic Density}
\label{sec:relic}
\noindent
For the considered DFs, the mass splitting can be parametrized as $\delta=m_2-m_1$. However, the DFs being almost degenerate, $\delta/m_1\ll 1$. To be specific, for $\delta\leq \mathcal{O}(100)$ keV and $m_1\approx m_2\sim\mathcal{O}(100)$ GeV, $\delta/m_1\leq \mathcal{O}(10^{-6})$ and the two DFs can be safely treated as a single fermionic field $\chi$ with mass $m_\chi=(m_1+m_2)/2$ at the freeze-out. In the early universe, a thermal equilibrium was maintained between the SM particles and $\chi$ through the scalar-mediated annihilation channels $\overline{\chi}\chi\leftrightarrow\overline{\rm SM}\,\,{\rm SM}$. After decoupling, the abundance of $\chi$ can be obtained by solving the Boltzmann equation:
\begin{align}
\frac{dn_\chi}{dt}+3\mathcal{H}n_\chi=-\left\langle\sigma_{\rm An}\mathtt{v}\right\rangle\Big[n_\chi^2-(n_\chi^{\rm eq})^2\Big]\,,
\label{eq:boltz}
\end{align}
where $n_\chi$ stands for the number density of $\chi$ with the superscript `eq' representing its equilibrium value. $\mathcal{H}$ denotes the Hubble parameter, and $\left\langle\sigma_{\rm An}\mathtt{v}\right\rangle$ is the thermal averaged annihilation cross section times the relative velocity~($\mathtt{v}$) of $\chi$. 
\begin{figure}[!ht]
\centering
\includegraphics[scale=0.68]{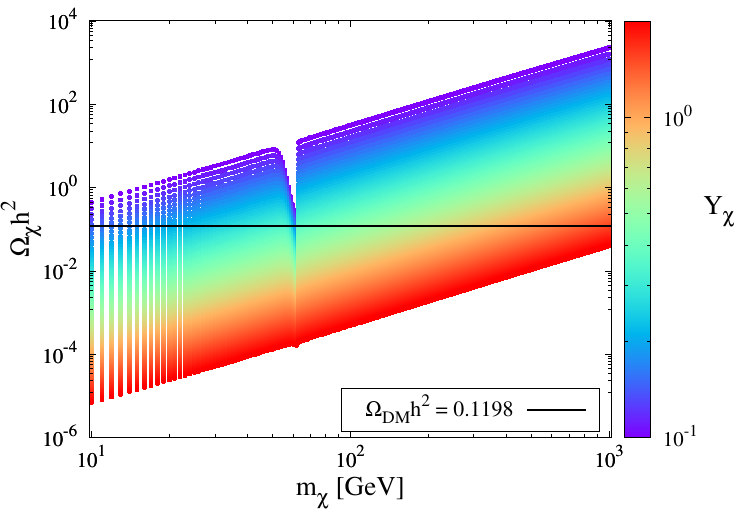}
\caption{Variation of relic density as a funtion of $m_\chi$ and $Y_\chi$. The black solid line represents the observed DM abundance from Planck~\cite{Planck:2018vyg}. $M_{h_1}$ has been fixed at 2 GeV for the analysis.}
\label{fig:relic} 
\end{figure}
Eq.~\eqref{eq:boltz} has been solved numerically using $\mathtt{micrOMEGAs\_7.1.4}$~\cite{Belanger:2026asz}. Fig.~\ref{fig:relic} depicts the variation of the relic density of $\chi$ as a function of $m_\chi$ and $Y_\chi$. Note that the upper bound on $Y_\chi$ is set by the perturbative unitarity: $|Y_\chi|<\sqrt{4\pi}$~\cite{Allwicher:2021rtd}. The resonance dip at $m_\chi=62.5$ GeV corresponds to $M_{h_2}=125$ GeV. One can see a similar dip at $m_\chi=1$ GeV, as the results displayed in Fig.~\ref{fig:relic} correspond to $M_{h_1}=2$ GeV. The black line marks the observed DM abundance reported by the Planck collaboration, i.e., $\Omega_{\rm DM}h^2=0.1198\pm 0.0012$~\cite{Planck:2018vyg}. Therefore, $\chi$ can produce the correct relic density over the entire considered range of $m_\chi$ with $Y_\chi\leq\mathcal{O}(1)$. 
\section{Inelastic DM-Nucleus Scattering}
\label{sec:scat}
\noindent
In the proposed framework, an inelastic DM-nucleus scattering originates from the $h_{1,\,2}$-mediated off-diagonal Yukawa couplings between the DFs. With the considered mass hierarchy, i.e., $m_2>m_1$ or $\delta>0$, $\chi_1$ up-scatters to $\chi_2$ resulting in an endothermic process: $\chi_1+N\to\chi_2+N$, where $N$ stands for the target nucleus. The threshold velocity~($\mathtt{v}^{\rm Th}$) to produce a nuclear recoil at energy $E_R$ can be defined as,
\begin{align}
\mathtt{v}^{\rm Th}(E_R)=\frac{1}{\sqrt{2\,\mathbb{M}_NE_R}}\left(\frac{\mathbb{M}_NE_R}{\mu_{\chi N}}+\delta\right)\,,
\end{align}
where, $\mathbb{M}_N$ is the mass of the target nucleus and $\mu_{\chi N}=m_1 \mathbb{M}_N/(m_1+\mathbb{M}_N)$ defines the reduced mass of the DM-nucleus 2-body system. At $E_R=\mu_{\chi N}\delta/\mathbb{M}_N$, $\mathtt{v}^{\rm Th}(E_R)$ reaches its absolute minimum, given by,
\begin{align}
\mathtt{v}^{\rm Th}_{\rm min}=\sqrt{\frac{2\delta}{\mu_{\chi N}}}\,.
\end{align}
Elastic scattering corresponds to $\delta=0$. However, for the single nuclear recoil at $E_R=248$ keV, inelastic DM scatterings are particularly favoured, as the non-zero mass splitting sets a threshold for the DM-nucleus scattering and one can trivially explain the non-observation of similar recoil signals at lower energy values. 

Considering the LZ detector, the differential event rate per unit detector mass for the DM-Xe scattering can be cast as~\cite{Tucker-Smith:2001myb},
\begin{align}
\frac{dR}{dE_R}=\mathbb{N}_{\rm Xe}\,\frac{\rho_\chi}{m_1}\int_{\mathtt{v}>\mathtt{v}^{\rm Th}_{\rm min}}f_{\rm Gal}(\vec{\mathtt{v}}+\vec{v}_e)\left(\frac{d\sigma}{dE_R}\right)\mathtt{v}\,\, d^3\mathtt{v}\,,
\label{eq:rate}
\end{align} 
where, $\mathbb{N}_{\rm Xe}$ is the number density of Xe nucleus. For $^{131}$Xe, $\mathbb{M}_N\approx 122$ GeV. $\rho_\chi=0.3$ GeV/cm$^3$ defines the local DM density, with $\vec{\mathtt{v}}$ standing for the velocity of the incoming DM state with respect to the Earth. Assuming the isothermal Standard Halo Model~(SHM), the DM velocity distribution in the galactic frame can be defined as,
\begin{align}
f_{\rm Gal}(\vec{\mathtt{v}})\propto\exp(-\mathtt{v}^2/v_0^2)\Theta(v_{\rm esc}-\mathtt{v})\,,
\end{align}
with $\Theta$ representing the Heaviside step function. The distribution shows a hard cut-off at the galactic escape velocity $v_{\rm esc}= 550$ km/s with a velocity dispersion, $v_0 = 220$ km/s. The velocity of earth at the galactic rest frame is given by $\vec{v}_e$. The differential DM-nucleus scattering cross section, $d\sigma/dE_R$, can be defined as,
\begin{align}
\frac{d\sigma}{dE_R}=\frac{\mathbb{M}_N\sigma_n}{2\mu_R^2\mathtt{v}^2}\left[Z(f_p/f_n)+(A-Z)\right]^2\times\mathcal{F}^2(E_R)\,,
\end{align}
Here $\sigma_n$ denotes the DM-neutron reference cross section at zero momentum transfer and $\mu_R$ is the $\chi_1$-nucleon reduced mass. $A$ and $Z$ denote the mass number and atomic number of the detector nucleus, respectively. $f_p$ and $f_n$ are the effective DM-proton and DM-neutron couplings, while the nuclear structure effects can be parametrized through the Helm form factor as~\cite{Engel:1991wq},
\begin{align}
\mathcal{F}^2(E_R)=\left(\frac{3j_1(qr_0)}{qr_0}\right)^2\,e^{-s^2q^2}\,,
\end{align} 
where, $j_1(x)$ is the spherical Bessel function of first of kind, $q=\left(2\mathbb{M}_NE_R\right)^{1/2}$, $s\simeq 0.9$ fm, and $r_0=\left[\ell^2+(7/3)\pi^2\beta^2-5s^2\right]^{1/2}$, with $\beta\simeq 0.52$ fm, and $\ell=1.23A^{1/3}-0.6$. For the DFs, 
\begin{align}
\sigma_n=\frac{\mu_R^2}{\pi}|f_n|^2\,.
\end{align}
In the present model, the effective DM-nucleon couplings can be formulated as,
\begin{align}
f_{a}=M_a\left[\sum\limits_{q=u,d,s}\frac{\xi_q}{m_q} f_q^{(a)}+\frac{2}{27}f_G^{(a)}\sum\limits_{q=c,b,t}\frac{C_q\xi_q}{m_q}\right]\,.
\label{eq:fa}
\end{align} 
Here, $a$ stands for the nucleons, i.e., either $n$ or $p$, while $M_a$ and $m_q$ represent the nucleon and quark masses, respectively. The form factors $f_{u,d,s}^{(a)}$ can be determined through the lattice QCD calculations as~\cite{Thomas:2012tg,Belanger:2013oya}\,\footnote{Slightly different values can be obtained by using chiral perturbation theory~\cite{Alarcon:2011zs,Crivellin:2013ipa}.}, 
\begin{align}
&f^{(p)}_{u} =0.0153,\quad f^{(p)}_{d} =0.0191,\quad f^{(p)}_{s} =0.0447, \nonumber\\
&f^{(n)}_{u} =0.0110,\quad f^{(n)}_{d} =0.0273,\quad f^{(n)}_{s} =0.0447\,,
\label{eq:ff}
\end{align}
whereas, for the heavy quarks, 
\begin{align}
f_{c,b,t}^{(a)}=\frac{2}{27}f_G^{(a)}=\frac{2}{27}\left(1-\sum\limits_{q=u,d,s}f_q^{(a)}\right)\,.
\end{align}
However, ignoring the differences between nucleons, one obtains $f^{(a)}_G \simeq 0.92$.  $C_q=1+11\alpha_s(m_q)/{4\pi}$ is the leading order QCD correction for the heavy quarks, with $\alpha_s$ denoting the strong coupling constant. Finally, the DF-quark effective coupling is given by,
\begin{align}
&\xi_q=Y_\chi\,y_q\left(\frac{1}{M_{h_1}^2}-\frac{1}{M_{h_2}^2}\right)\sin\theta\cos\theta\,,
\end{align} 
where, $y_q$ represents the quark Yukawa coupling with $y_q^{\rm SM}=\sqrt{2}\,m_q/v$. However, it is well known that the fermion Yukawa couplings can deviate from their SM values in the presence of a higher dimensional effective operator~\cite{Das:2020ozo,Allwicher:2025mmc}. In the SM effective field theory~(SMEFT) Warsaw basis~\cite{Grzadkowski:2010es}, there is just one dimension-6 effective operator that results in flavor-specific Yukawa modifications unsuppressed by the corresponding fermion masses: $\mathbf{O}_{fH}=\left(H^\dagger H\right)\overline{F}_LH f_R$. For a given fermion generation, $F_L$ and $f_R$ represent the left-chiral $SU(2)_L$ doublet and right-chiral $SU(2)_L$ singlet, respectively. Let's assume a case where only the down-quark Yukawa coupling gets modified through $\mathbf{O}_{dH}$, i.e.,
\begin{align}
-\mathcal{L}^{\,d}_{\rm Yukawa}=y_d^{\rm SM}\,\,\overline{Q}_LH d_R-\frac{\mathcal{Y}_d}{\Lambda^2}\left(H^\dagger H\right)\overline{Q}_LH d_R+{\rm h.c.}\,.
\label{eq:dNP}
\end{align}  
Here $\mathcal{Y}_d$ denotes the Wilson coefficient determined by the details of the NP model. Therefore, after EWSB, one gets the physical $d$-quark mass and Yukawa coupling as,
\begin{align}
m_d=&~\left(\frac{y_d^{\rm SM}}{\sqrt{2}}-\frac{\mathcal{Y}_d\epsilon}{2\sqrt{2}}\right)v\,,\nonumber\\
y_d=&~\left(y_d^{\rm SM}-\frac{3\mathcal{Y}_d\epsilon}{2}\right)=\frac{\sqrt{2}\, m_d}{v}-\mathcal{Y}_d\epsilon\,,
\end{align}
where $\epsilon=(v/\Lambda)^2$. Though most of the experimental constraints are insensitive to the sign of the quark Yukawa couplings, Ref.~\cite{Bonner:2016sdg} shows that using the high-luminosity run of the Large Hadron Collider~(HL-LHC), one can restrict $y_d/y_d^{\rm SM}\in(-800,\,300)$. Note that negative quark Yukawa couplings are phenomenologically significant, particularly for the DM direct detection~\cite{Das:2020ozo}. Moreover, the negative non-SM-like values lead to an upper bound on the NP scale. One can trivially check that with $\mathcal{Y}_d\sim\mathcal{O}(1)$ and $m_d=m_d^{\rm SM}=4.7$ MeV, a negative $y_d$ results in $\Lambda<47.4$ TeV. Thus, by varying the $d$-quark Yukawa coupling, one can alter the effective DM-nucleon interaction strength $f_a$ without affecting Eq.~\eqref{eq:Yx}. Moreover, the impact of a non-SM-like $d$-quark Yukawa coupling on $\Omega_\chi h^2$ is negligible. 

Eq.~\eqref{eq:rate} has been evaluated using the publicly available Python package $\mathtt{WimPyDD\_2.0.4}$~\cite{Jeong:2021bpl}, which encodes the DM-nucleon effective interaction through two parameters: the isospin-violation parameter $r=f_n/f_p$ and an effective mass scale $\mathcal{M}=1/\sqrt{f_p}$. Table~\ref{tab:BP} enlists two particular cases for a thermal DM candidate at $m_1=1$ TeV. 
\begin{table}[!ht]
\centering
\begin{tabular}{|c|c|c|}
\hline
{\bf Parameters} & {\bf Case-I} & {\bf Case-II}\\
\hline\hline
$m_1$ [GeV] & $10^3$ & $10^3$\\
\hline
$\delta$ [keV] & 350 & 340\\
\hline
$Y_\chi$ & 1.49 & 1.49\\
\hline
$M_{h_1}$ [GeV] & 2 & 0.9\\
\hline
$\Omega_\chi h^2$ & 0.12 & 0.12\\
\hline
$y_d/y_d^{\rm SM}$ & 1 & $-11.77$\\
\hline
$\mathcal{M}=1/\sqrt{f_p}$ ~[GeV] & 181.42 & 184.56\\
\hline  
$r$ & $1.01$ & $-\,0.7$\\
\hline
$\sigma_n$ $[\,{\rm cm}^2\,]$ & $1.03\times 10^{-37}$ & $4.60\times 10^{-38}$\\
\hline
\end{tabular}
\caption{Parameters corresponding to the two specific cases. Case-I preserves the SM Yukawa interaction, whereas Case-II corresponds to a non-SM-like negative $d$-quark Yukawa coupling. As before, $\sin\theta=0.05$ has been considered for the computation.}
\label{tab:BP}
\end{table}
Case-I stands for the SM-like situation where all the quark Yukawa couplings are fixed at their SM values, whereas Case-II represents the scenario where $y_d=-11.77y_d^{\rm SM}$, leading to $r=-0.7$. The latter corresponds to a significant suppression in the DM-Xe scattering cross section due to isospin violation~\cite{Feng:2011vu}. However, its implications for the differential event rate can be interesting.
\begin{figure}[!ht]
\centering
\includegraphics[scale=0.68]{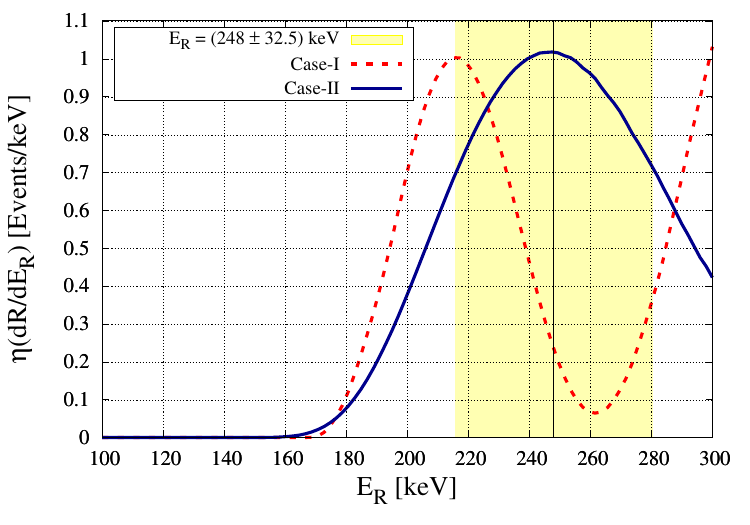}
\caption{Variation of the normalized differential event rate as a function of the recoil energy within the window $E_R\in[100,\,300]$ keV. $\eta=2.84$ tonne-years defines the exposure in the LZ experiment. The shaded region represents the observed LZ event at $E_R=248\pm 32.5$ keV, with the vertical black line marking the central value.}
\label{fig:rate}
\end{figure}
Fig.~\ref{fig:rate} shows the variation of differential event rate as a function of the recoil energy, $E_R$, where the red dashed line marks Case-I and the dark-blue line represents Case-II, respectively. The shaded region depicts $E_R=248\pm 32.5$ keV, with the black vertical line showing the central value at 248 keV. Note that the event rate has been normalized with an exposure of $\eta=2.84$ tonne-years $\approx 1.037\times 10^6$ kg-days. For Case-I, the differential event rate shows a peak around $E_R\sim 216$ keV, while for Case-II, $dR/dE_R$ becomes maximum at $E_R\sim 248$ keV, indicating exact agreement with LZ230616. Note that in Case-II, the values of $M_{h_1}$ and $\delta$ have been slightly altered to make the differential event rate comparable to that of Case-I; the alteration has no visible impact on the position of the peak. 

For the observed LZ event, the two-sided $90\%$ confidence-level~(CL) bound on the DM-nucleon interaction corresponding to a 1 TeV scalar-portal DM with $\delta=340$ keV can be approximately read as $5\times 10^{-41}~{\rm cm}^2<\sigma_n< 10^{-39}$ cm$^2$~\cite{LZ:2026axp,Freese:2026sga}. Clearly, the Case-II doesn't comply with this bound. However, the LZ analysis follows from the assumption that the SM quark sector is free from any NP correction. As already mentioned, the particular choice $y_d=-11.77y_d^{\rm SM}$ results in a remarkable suppression in the DM-Xe scattering cross section. Thus, in the presence of a quarkophilic NP, LZ230616 can be explained even with a comparatively large $\sigma_n$.
\section{Conclusion}
\label{sec:conc}
\noindent
The proposed model has augmented the SM particle specturm with two nearly degenerate VLFs $\chi_1$, $\chi_2$~(the DFs), and a scalar $\Phi$. The DFs are protected by an unbroken $Z_2$ symmetry, with a mass hierarchy $m_2>m_1$. Moreover, all the BSM states are charged under a local $U(1)^\prime$, such that $\Phi$ can have only an off-diagonal Yukawa coupling with the DFs. However, the $U(1)^\prime$ has been softly broken through a tiny mass dimensional coupling $\mu$, resulting in a mixing between $\phi$ and the SM Higgs, $h$. Though the soft $U(1)^\prime$ breaking can also induce a mixing between the DFs, leading to diagonal Yukawa couplings as well in the physical basis, the mixing can be ignored in the limit $\mu\ll v$. The model presents a simple framework where the lighter DF can up-scatter via a scalar mediator, with the effective DM-nucleus interaction being a function of the quark Yukawa couplings. Considering the DFs to be completely degenerate at decoupling, the observed DM abundance can be explained for $m_1\approx m_\chi\in [10,\,1000]$ GeV with $Y_\chi\leq\mathcal{O}(1)$. Further, in the context of LZ230616, the paper considers two cases --- $(i)$ all the quark Yukawa couplings are at their SM values, $(ii)$ only the down quark Yukawa coupling can assume a negative value. The second case can be realized with a dimension-6 effective operator at a NP scale $\Lambda<47.4$ TeV. With $y_d=-11.77y_d^{\rm SM}$, the isospin violation parameter $r$ charges from $1.01$~(approximately isoscalar) to $-0.7$. The differential event rate shows a maximum at 216 keV for the first case, whereas with the aforementioned negative $y_d$ value, the peak shifts to 248 keV, showing excellent agreement with the LZ recoil event. The result stems from a suppression in the DM-Xe scattering cross section due to the $\mathbf{O}_{dH}$-induced isospin violation. Therefore, LZ230616 may not only be a hint of non-gravitational DM-SM interaction, but may also encapsulate the signature of a NP interaction in the quark sector. However, the possibility of a quarkophilic NP can only be tested through future colliders.

%%%%%%%%%%%%%%%%%%%%%%%%%%%%%%%%%%%%%%%%%%%%%%%%%%%%%%%%%%%%%%%%%%%%%%%%%%%%%%%%%%%%%%    
\bigskip
\small \bibliography{LZ_DM}{}
\bibliographystyle{JHEPCust}    
    
\end{document}